\documentclass[letterpaper,10pt,notitlepage]{article}
\pdfoutput=1
\usepackage{jheppub} 

\usepackage{mathrsfs}
\usepackage{color}
\usepackage{graphicx}
\usepackage{amsmath}
\usepackage{amssymb}
\usepackage{bbm}

\def\Of{{O(k^4)}}
\def\L{L}

\def\ti{t}
\def\bi{{\mathbbm i}}

\def\Ph{{\cal{\phi}}}
\def\tph{{\theta^\phi}}
\def\dzp{\delta \zeta_c}
\def\dzz{\delta \zeta_0}
\def\djp{\delta j_c}
\def\djz{\delta j_0}

\def\Ft{\tilde{\xi}}
\def\La{{^{_\Lambda}}}
\def\cc{{_{c}}}
\def\zz{{_0}}
\def\sfl{s_\cc} 
\def\Tf{T_\cc}
\def\tsc{\tilde{s}_\cc} 
\def\tTc{\tilde{T}_\cc}

\def\lf{\lambda_\cc}
\def\kf{\kappa_\cc}

\def\cgL{c_{\gamma}}
\def\cgz{c_{\zz}}
\def\cp{c_\La} 
\def\ekf{\ell_\cc}
\def\tauo{\tau_c}
\def\bto{{\bar{\tau}_c}^{-1}}
\def\tka{\tilde{\kappa}}

\def\rz{r_\zz}
\def\rh{r_h}
\def\rc{r_\cc}

\def\cz{c_\phi}
\def\ekz{\ell_\zz}
\def\lz{\lambda_\zz}
\def\kz{\kappa_\zz}
\def\sz{s_\zz}
\def\Tz{T_\zz}
\def\tTz{\tilde{T}_\zz}
\def\tsz{\tilde{s}_\zz}

\def\tauz{\tau_\zz }
\def\btz{{\tilde{\tau}_\zz}^{-1}}

\def\DC{_{DC}}

\def\pc{\mathbbm{p}}
\def\ec{\mathbbm{e}}
\def\tpc{\tilde{\pc}}
\def\tec{\tilde{\ec}}
\def\tf{\tilde{f}}

\def\cE{{\cal{E}}}
\def\cP{{\cal{P}}}
\def\tcE{\tilde{\cE}}
\def\tcP{\tilde{\cP}}

\def\cO{{\cal{O}}}
\def\cQ{{\cal{Q}}}
\def\cT{{\cal{T}}}
\def\cA{{\cal{I}}}
\def\OA{{\langle O_\cA\rangle}}
\def\tr{{\tilde{r}}}
\def\trp{{\bar{r}}}

\def\m{\mu}
\def\n{\nu}
\def\g{{\gamma}}
\def\s{\sigma}
\def\p{\partial}
\def\d{\mathrm{d}}

\def\de{\delta}

\def\ph{{\phi}}

\def\be{\begin{equation}}
\def\ee{\end{equation}}
\def\beq{\begin{eqnarray}}
\def\eeq{\end{eqnarray}}
\def\bea{\begin{eqnarray}}
\def\eea{\end{eqnarray}}
\def\nn{{\nonumber}}
\renewcommand{\(}{\left(}
\renewcommand{\)}{\right)}
\renewcommand{\[}{\left[}
\renewcommand{\]}{\right]}

\def\cd{\color[rgb]{1.00,0.00,0.00}} 

\title{Rindler Fluid with Weak Momentum Relaxation} 

\author[a]{Sunly Khimphun,}
\author[a]{Bum-Hoon Lee,}
\author[b, c]{Chanyong Park,}
\author[a, b]{Yun-Long Zhang}

\affiliation[a]{Center for Quantum Spacetime(CQUeST) and Department of Physics, Sogang University,\\
 35 Baekbeom-ro, Sinsu-dong, Mapo-gu, 121-742, Seoul, Korea}
\affiliation[b]{Asia Pacific Center for Theoretical Physics, APCTP Headquarters,\\
 67 Cheongam-ro, Hyoja-dong, Nam-gu, 790-784, Pohang, Korea}
\affiliation[c]{Department of Physics, Pohang University of Science and Technology(POSTECH),\\
 77 Cheongam-ro, Jigok-dong, Nam-gu, 790-784, Pohang, Korea}

\emailAdd{kpslourk@sogang.ac.kr}
\emailAdd{bhl@sogang.ac.kr}
\emailAdd{chanyong.park@apctp.org}
\emailAdd{yunlong.zhang@apctp.org}

\abstract{
We realize the weak momentum relaxation in Rindler fluid, which lives on the time-like cutoff surface in an accelerating frame of flat spacetime. The translational invariance is broken by massless scalar fields with weak strength. Both of the Ward identity and the momentum relaxation rate of Rindler fluid are obtained, with higher order correction in terms of the strength of momentum relaxation. The Rindler fluid with momentum relaxation could also be approached through the near horizon limit of cutoff AdS fluid with momentum relaxation,  which lives on a finite time-like cutoff surface in Anti-de Sitter(AdS) spacetime, and further could be connected with the holographic conformal fluid living on AdS boundary at infinity. Thus, in the holographic Wilson renormalization group flow of the fluid/gravity correspondence with momentum relaxation, the Rindler fluid can be considered as the Infrared Radiation(IR) fixed point, and the holographic conformal fluid plays the role of the ultraviolet(UV) fixed point.

}

\arxivnumber{1705.05078}

\preprint{http://dx.doi.org/10.1007/JHEP01(2018)058}
\dedicated{January 18, 2018}

\begin{document}

\maketitle


\vspace{-15pt}
\section{Introduction}
There are several  approaches to describe the duality between gravity and lower dimensional fluid,
such as  the membrane paradigm~\cite{Damour:1978cg,Price:1986yy}, and the holographic hydrodynamics based on the AdS/CFT correspondence~\cite{Aharony:1999ti,Policastro:2001yc,Policastro:2002se,Policastro:2002tn}.
In the long wavelength and low frequency limit,
the holographic hydrodynamics is quite  universal, which can be considered either on the stretched horizon \cite{Kovtun:2003wp},
or on a finite membrane which can flow from the horizon to the boundary  \cite{Iqbal:2008by}.
Another systematic way to study the duality  is the fluid/gravity correspondence~\cite{Bhattacharyya:2008jc} which has been further explored in \cite{VanRaamsdonk:2008fp,Bhattacharyya:2008ji,Haack:2008cp,Bhattacharyya:2008mz,Erdmenger:2008rm,Banerjee:2008th,Hur:2008tq,Bhattacharyya:2008kq,Kim:2013wiz,Ashok:2013jda}. 
This approach can also be applied to the  membrane paradigm to analyze the horizon dynamics \cite{Eling:2009pb,Eling:2009sj}.

To build up the connection between the holographic hydrodynamics in AdS/CFT correspondence and membrane paradigm near a black hole horizon, a finite time-like cutoff surface can be introduced in the AdS spacetime~\cite{Bredberg:2010ky}.
The dual fluid lives on the cutoff surface with a Dirichlet boundary condition as well as a regular horizon,
and the cutoff can be taken near the boundary or the horizon separately.
This approach has been further explored in~\cite{Cai:2011xv,Kuperstein:2011fn,Brattan:2011my,Niu:2011gu,Eling:2011ct,Cai:2012vr,Bai:2012ci,Cai:2012mg,Zou:2013ix,Emparan:2013ila,Kuperstein:2013hqa,Pinzani-Fokeeva:2014cka}.
In this case, the asymptotically AdS boundary is not necessary anymore, and one should consider  the perturbations between the horizon and cutoff surface.

One significant example of this approach is the so-called Rindler fluid, which lives on the time-like cutoff surface in accelerating frame of flat spacetime \cite{Bredberg:2011jq}. It has been developed  systematically with the fluid/gravity duality in derivative expansion~\cite{Compere:2011dx,Chirco:2011ex,Compere:2012mt,Eling:2012ni,Eling:2012xa,Meyer:2013sva},
and an interesting recursive relation in Rindler fluid is explored in \cite{Lysov:2011xx,Huang:2011he,Cai:2013uye,Cai:2014ywa,Cai:2014sua,Hao:2014xva}.
Rindler fluid can be approached through the near horizon limit of the dual fluid on the cutoff surface in AdS  \cite{Matsuo:2012pi},
which will be named as ``cutoff AdS fluid" in this paper.
It can also be  related to the holographic fluid on the boundary of AdS through AdS/Rindler correspondence \cite{Caldarelli:2012hy,Caldarelli:2013aaa}.
With a Dirichlet cutoff surface outside the horizon, the causal structure of the  Rindler spacetime with a cutoff is similar to the Poincare patch of AdS spacetime,
which provides one frame to study the holography in flat spacetime and motivate us to study the Rindler fluid in more details.

An advantage of AdS/CFT is the convenience to study  transport properties  in strongly coupled systems \cite{Herzog:2007ij,Hartnoll:2007ih}.
In addition, plentiful progress has been made in the holographic models with broken translational invariance 
\cite{Horowitz:2012ky,Horowitz:2012gs,Vegh:2013sk,Davison:2013jba,Blake:2013bqa,Blake:2013owa,Donos:2013eha}.
In one widely used holographic model of Einstein-Maxwell theory, the momentum relaxation is caused by spatially dependent massless neutral scalars \cite{Andrade:2013gsa,Davison:2014lua,Davison:2015bea}.
Based on this simple model, the hydrodynamical description of the weak momentum relaxation is realized in the fluid/gravity correspondence   \cite{Blake:2015epa,Blake:2015hxa},
which can reproduce the known features of holographic transport coefficients. 
Here weak means the small strength of momentum relaxation $k$, and the hydrodynamical derivative expansion parameter is of order $k^2$.
Both of the Ward identity and momentum relaxation rate can be obtained in the expansion of small parameter $k$,
and higher order corrections appear beyond the hydrodynamical assumptions of momentum relaxation in \cite{Hartnoll:2007ih}.
Further studies of the holographic hydrodynamics without translational symmetry can also be found in \cite{Hartnoll:2016tri,Alberte:2016xja,Burikham:2016roo}.

On the other hand, the holographic transport coefficients are always expressed in terms of the horizon data,
and can be obtained by solving the hydrodynamical equations on black hole horizon \cite{Donos:2014cya,Donos:2015gia,Banks:2015wha}.
Since Rindler frame is quite universal and a good approximation to describe the near horizon limit of  non-extremal black holes,
it is interesting to see whether we can realize the similar momentum relaxation in Rindler fluid.
In this paper, we will start with the relativistic neutral Rindler fluid in arbitrary dimensions.
As in \cite{Blake:2015epa}, we consider  the strength of the momentum relaxation $k$ as the small parameter, 
and assume the derivative expansion in terms of coordinates on the cutoff surface that $\p_a= (\p_t, \p_i) \sim k^2$.
After solving the field equations up to {order $k^3$} with appropriate boundary conditions,
we can read off the Ward identity, momentum relaxation rate, and heat conductivity of Rindler fluid.

In order to confirm our results, as well as to build the relation between  Rindler fluid and the dual boundary fluid in AdS,
we also study the momentum relaxation from fluid/gravity duality on the Dirichlet cutoff surface in AdS.
Notice that the forced fluid dynamics dual to AdS gravity with massless scalar fields has already been studied on the cutoff surface in \cite{Cai:2012vr}, as well as the generalization in arbitrary  dimensions on the boundary~\cite{Ashok:2013jda},
which are helpful for us to compare the results. 
For convenience, we will use the name ``cutoff AdS fluid" to indicate the ``fluid dual to AdS spacetime with a finite cutoff surface''.
Since the calculation of fluid/gravity duality on the cutoff surface is more complicated than that on the boundary,
we turn off the Maxwell field in the paper,
which does not affect our main purposes to extract the Ward identity and momentum relaxation rate. 

In all calculations, we will impose the Dirichlet boundary condition at the cutoff surface,  and require the regular boundary condition at the horizon. 
We will show that the Rindler fluid with momentum relaxation could also be approached through near horizon limit of the ``cutoff AdS fluid"  with momentum relaxation,  which lives on a finite time-like cutoff surface in  Anti-de Sitter(AdS) spacetime, and further could be connected with the holographic conformal fluid living on AdS boundary at infinity.
From the viewpoint of holographic Wilson RG flow in the fluid/gravity duality with momentum relaxation, the Rindler fluid can be considered as the IR fixed point, and the holographic conformal fluid plays the role as the UV fixed point.

In the following section \ref{RindlerMR}, we  study the Rindler fluid with weak momentum relaxation perturbatively.
In section \ref{AdSCut}, we study the  cutoff AdS fluid with weak momentum relaxation, 
and analyze both of the near horizon limit and near boundary limit,
which build the flow from Rindelr fluid to holographic  fluid on the AdS boundary.
In section \ref{RGflow}, we study the relation between Rindler fluid and holographic Wilson RG flow in fluid/gravity duality with the momentum relaxation.
In section \ref{conclusion}, we make the summary and discussion of further topics.

\section{Momentum Relaxation in Rindler Fluid}
\label{RindlerMR}

We start with the Einstein-Hilbert action in $p+2$ dimensional flat spacetime, with $p$ massless scalar fields $\ph_\cA$ and
${  \cA}=1,2,...,p$, 
\begin{align}
S_0=\frac{1}{16\pi G_{p+2}}\int \d^{p+2} x\sqrt{-g} \[R -\frac{1}{2}\sum_{\cA=1}^{p}  (\p \ph_\cA)^2 \]
-\frac{1}{8\pi G_{p+2}}\int \d^{p+1} x\sqrt{-\gamma} K.
\end{align}
We use $\mu,\nu=0,1,...,p+1$ to indicate the indexes of bulk spacetime.
Varying the action with respect to the metric $g_{\mu\nu}$ yields the gravitational and scalar field equations,
\begin{align}
R_{\mu\nu}-\frac{1}{2}R g_{\mu\nu} &=\frac{1}{2} \sum_{\cA=1}^p\[\p_\m\ph_\cA \p_\n\ph_\cA-\frac{1}{2} g_{\m\n}(\p \ph_\cA)^2\],\label{Reomg}\\
\nabla^2\ph_\cA &= 0,\qquad \cA=1,2,...,p\,. \label{Reomp}
\end{align}

If we consider $k$ as a small parameter,
the following $p+2$ dimensional Rindler metric with corrections of order $k^2$,
is a perturbed solution of the field equations \eqref{Reomg} up to $k^2$, 
\begin{align}\label{Rindler}
\d s_{p+2}^2 &=- 2\kz { (r-\rz)}  \d t^2+2\d t \d r+\delta_{ij}\d x^i \d x^j\nn\\
&\quad - \frac{p }{4} (r-\rz)(r-\rc) k^2 \d\ti^2 - \frac{(r-\rc)}{2\kz} k^2\delta_{ij}\d x^i \d x^j+\Of,\\
 \ph_\cA&=  k\, x_i  {\delta^i}_\cA,\qquad  x_\cA= x_1,x_2,...,x_p\, . 
\end{align}
Here $\kz$ is a constant to indicate the surface gravity in the accelerating frame.
We have chosen the condition that the horizon is located at $r=\rz$ even with the metric corrections of $k^2$.
 Also, on the timelike hypersurface $\Sigma_c$ with $r=\rc$, the induced metric is intrinsic flat.

%


\subsection{Fluid/Gravity Duality in Rindler Spacetime with cutoff} 
In order to study the fluid dual to Rindler spacetime with momentum relaxation,
we make the following coordinate transformation
\begin{align}
(r -\rz) \to \frac{\lz^2}{2\kz}+(r -\rc) ,\qquad 
\ti\to \frac{1}{\lz}\ti,\qquad 
\lz\equiv \sqrt{{2\kz}{(\rc-\rz)}}.
\end{align}
$\lz$ is a rescale of the time coordinate.
The Rindler metric in $p+2$ dimension, which is just the first line in \eqref{Rindler}, becomes
\begin{align}\label{Rmetric0}
\d s_{p+2}^2 
&=-\[1+ \frac{2\kz}{\lz^2} (r-\rc)\]\d\ti^2+    \frac{2}{\lz}  \d\ti \d r+\delta_{ij}\d x^i \d x^j.
\end{align}
One can read off the horizon position through
\begin{align} \label{Rrz}
 g_{tt}(\rz)=0~~\Rightarrow~~ 
 \rz=\rc-\frac{\lz^2}{2\kz}.
 \end{align}
 It will become clear that if we introduce the notation ${{\pc}} = {2\kz}/\lz$ and set $2\kz =1$, we can recover the set up in \cite{Compere:2012mt}.
The $p+1$ dimensional induced metric on the  timelike hypersurface $r=\rc$ is
\begin{align}\label{RindlerCut}
\d s_{p+1}^2=\gamma_{ab}\d x^a \d x^b=\eta_{ab}\d x^a \d x^b=-  \d\ti^2+\delta_{ij}\d x^i \d x^j~,
\end{align}
with $a,b=0,1,...,p$ .

As the usual set up in Rindler fluid,
we need to make the boost transformation associated with the hypersurface $\d t\to -u_a \d x^a, \delta_{ij}\d x^i \d x^j\to h_{ab}\d x^a \d x^b $, with  the projection tensor
$h_{ab}=\eta_{ab}+u_a u_b$. 
We  assume both of the $(p+1)$ velocity $u_a$ and the position of the horizon $\rz$ in \eqref{Rrz} to be $x^a$ dependent.
Then we can solve the field equations, \eqref{Reomg} and \eqref{Reomp},
in derivative expansion $\p_a\sim k^2$, with a small $k$.
We will use superscripts $^{(0)}, ^{(1)},...$ to indicate the number of derivatives.
Then, the metric and scalar field solutions of gravitational and scalar field equations \eqref{Reomg}   and  \eqref{Reomp} up to $O(k^3)$ turn out to be 
\begin{align}
\d {s}^2_{p+2} &={g}_{\mu\nu} \d x^\mu \d x^\nu= {g}_{ab} \d x^a \d x^b  - \frac{{2}}{\lz} u_a \d x^a \d r, \label{Rmetric}\\
{g}_{ab} &= {g}_{ab}^{{(0)}}+{g}_{ab}^{{(1)}} + O(k^4), \label{Rmetric1}\\
\ph_{\cA}&=\ph^{(0)}_{\cA}+\ph^{(1)}_{\cA}+\ph^{(2)}_{\cA} + \Of.   \label{Rscalar}
\end{align}
The leading order term of $g_{ab}$ and $\ph_{\cA}$ in derivative expansion are
\begin{align}
{g}_{ab}^{{(0)}}&= g_{uu}^{(0)}u_a u_b+h_{ab},\qquad h_{ab}\equiv\eta_{ab}+u_a u_b, \\
 \ph_\cA^{(0)}&= k\, x_i  {\delta^i}_\cA,\qquad\qquad ~~~ x_i=x_1,x_2,...,x_p\, . 
\end{align}
and the  $ g_{uu}^{(0)}$ is just the $g_{tt}$ component in the metric \eqref{Rmetric0},
\begin{align}
 g_{uu}^{(0)}= -1-\frac{2\kz}{\lz^2}(r-\rc)=-\frac{r-\rz}{\rc-\rz}.
\end{align}
Notice that  $\ph_\cA^{(0)} = k x_\cA$ is chosen as the laboratory frame where the velocity $u^a$ is defined.

The first order term of $g_{ab}$ in the derivative expansion \eqref{Rmetric1} is solved as
\begin{align}
{g}_{ab}^{{(1)}} 
&= g_{uu}^{(1)}(r) u_a u_b+2 u_{(a} g_{b)u}^{(1)}(r)  +g_{hh}^{(1)}(r)h_{ab}  +F_\s(r)\sigma_{ab} +F_{\ph}(r) \sigma^{\ph}_{ab}  ,  
\end{align}
where the metric components are given by
\begin{align}
g_{uu}^{(1)}(r)&=   \frac{(r-\rc)}{\lz} \[ \frac{2(p-1)}{p  } \theta  -2 D \ln {\lz}  + \Big(\frac{1}{4}\frac{r-\rc}{\rz-\rc  }  -\frac{p-1}{p}\Big) \frac{\lz}{2 \kz}   \theta^{\ph}  - {\dzz}  \theta^{\ph}\] , \label{Rguur} \nn\\ 
g_{hh}^{(1)}(r)&=- \frac{(r-\rc)}{2\kz} \frac{\theta^{\ph}}{p} ,  \nn\\
g_{au}^{(1)}(r)&= \frac{(r-\rc)}{\lz} \(a_a+ D_{a}^\bot \ln \lz  -\djz a_a^\ph \) , 
\end{align}
where $\delta\zeta_0$ and $\delta{j}_0$ are two integration constants when solving the Einstein equations.
The coefficients in front of shear tensors are determined to be
\begin{align}
F_{\sigma}(r)=0,\qquad
F_{\ph}(r)&=-\frac{(r-\rc)}{2\kz}.
\end{align}
Notice that $D \equiv u^a \p_a$,  $D_a^\bot \equiv h_a^b \p_b$ and the acceleration $a^a\equiv D u^a$  have been used, and the following notations  are defined through
\begin{align}
\theta^{\ph} &\equiv  \sum_{\cA=1}^p h^{ab}(D_a^\bot \ph_{\cA}^{(0)})(D_b^\bot \ph_{\cA}^{(0)}), \nn\\
a_a^{\ph} &\equiv   \sum_{\cA=1}^p (D_a^\bot \ph_{\cA}^{(0)})(D \ph_{\cA}^{(0)}),\nn\\
 \sigma^{\ph}_{ab} &\equiv  \sum_{\cA=1}^p (D_a^\bot \ph_{\cA}^{(0)})(D_b^\bot \ph_{\cA}^{(0)}) - \frac{1}{p}  \theta^{\ph} h_{ab}.
\end{align}
As in \cite{Blake:2015epa}, we ignore the terms $(D\ph_{\cA}^{(0)})^2$ which will not appear in the linear theory of our physical purpose.
Up to the second order in derivative expansion, the solutions of the scalar fields are
\begin{align}
\ph^{(1)}_{\cA} &= 0\,, \nn\\
\ph^{(2)}_\cA&=\frac{(r-\rc)}{2\kz} \[(D\ph_{\cA}^{(0)})\big( \frac{\lz}{4\kz} \theta^{\ph}-\theta\big)
+ (D_a^\bot\ph_{\cA}^{(0)})\( -2a_a+\djz a^\ph_a\)  \].
\end{align}
Since $\ph^{(2)}_\cA$ is non-trivial along the $r$ direction,
it behaves like the scalar hair in the Rindler spacetime with a Dirichlet boundary condition at the cutoff surface.

{\bf Dual Hydrodynamics.} ---
Now we will define the dual stress tensor and scalar operators on the timelike hypersurface $r=\rc$,
\begin{align}
\langle T_{ab} \rangle&= - \[ 2\({K}_{ab}-K {\g}_{ab} \)  + \frac{2p}{\rc}{\cgz} {\g}_{ab}+ \frac{\cz}{p-1} \frac{\lz}{2\kz}  \sum_{\cA=1}^p\((\p_a\ph_\cA\p_b\ph_\cA)-\frac{1}{2}{\g}_{ab} (\p\ph_\cA)^2\)\]\Big|_{r=\rc} ,\\
\OA&=-  \[n^{{\m}} \p_{{\m}} \ph_\cA+\frac{\cz}{p-1} \frac{\lz}{2\kz}  \p^2\ph_\cA\]\Big|_{r=\rc} .
\end{align}
Here ${K}_{ab}$ is the extrinsic curvature of the hypersurface, and we set $16\pi G_{p+2}=1$ for convenience.
Although on the cutoff surface the counter term is not necessary, we keep the undetermined constants here,
which are helpful to compare with the near horizon limit of the ``cutoff AdS fluid'' in section  \ref{AdSCut}.
Up to the first order in derivative expansion, we obtain
\begin{align}
\langle T_{ab} \rangle &=  (\ec + \zeta_{\zz}' D\ln\lz- 2\theta) \, u_a u_b+ \pc h_{ab}-2\eta_{\zz}\, \sigma_{ab}\nn\\ 
&\quad + \zeta'_\ph\theta^\ph u_a u_b+\zeta_\ph\theta^\ph h_{ab} + 2 j_\ph a^\ph_{(a} u_{b)} -2\eta_\ph \sigma_{ab}^\ph+\Of,\label{RTab}\\
\OA&= - (D \ph_\cA^{(0)}) -  
 \frac{1}{\pc} (D \ph_\cA^{(0)}) \big( \frac{ \theta^{\ph}}{2\pc} -\theta\big) 
+\frac{1}{\pc}  (D_a^\bot\ph_{\cA}^{(0)}) ( 2a_b-\djz a^\ph_b ) \eta^{ab} +\Of.
\end{align}
We have introduced the following notations of the background metric,
where the energy density ${\ec} $, pressure ${\pc}$, shear viscosity $\eta$, and effective bulk viscosity $\zeta'$ are
\begin{align}
{\ec} &= \frac{2p}{\rc}{\cgz}    ,\qquad
{\pc} =\frac{2\kz}{\lz} - {\ec} ,\qquad
\eta_{\zz}=1,\quad \zeta_{\zz}'=0. \label{Rbackground}
\end{align}
The coefficients before the terms from the scalar fields  are
\begin{align}
\eta_\ph &=\frac{1}{2\pc}\[1 +\frac{ {\cz}  }{ (p-1)}\],\qquad\qquad j_\ph  =\djz+ \frac{  {\cz}}{p-1}, \nn\\
\zeta_\ph &={{\dzz}}+ \frac{1}{ \pc} \frac{( p-2 )}{ 2p(p-1) } {\cz},\quad\qquad \zeta'_\ph= \frac{1}{ \pc}\[1 -\frac{ {\cz} }{ 2(p-1) }\].
\end{align}
On the other hand, the conservation equation of the stress tensor
$\p_a \langle {T^a}_{b} \rangle {=}\OA \p_b\ph_\cA$
 leads to
\begin{align}
(\ec_\zeta+\pc_\zeta) \theta+D ({\ec}_\zeta)&=  \sum_{\cA=1}^p  (D \ph_\cA^{(0)}) \OA+\cdots , \label{Rnst}\\
(\ec_\zeta+\pc_\zeta)  a_a+ D_a^\bot (\pc_\zeta) &=  \sum_{\cA=1}^p  (D_a^\bot \ph_{\cA}^{(0)}) \OA+\cdots  , \label{Rast}
\end{align}
where we have introduced the following notations
\begin{align}
\ec_\zeta &\equiv  \zeta'_\ph \theta^{\ph} ,\qquad 
\pc_\zeta \equiv\pc+ \zeta_\ph \theta^{\ph} . 
\end{align}
From  \eqref{Rnst}, we conclude that $\theta\sim O(k^4)$.
From \eqref{Rbackground}, we see that on the hypersurface $r=\rc$ the Smarr relation
of the thermodynamical quantities associated with the metric \eqref{Rmetric0} are still satisfied 
\begin{align}
{\ec} +{\pc} = \Tz  {\sz},\qquad   \Tz=\frac{\kz}{2 \pi \lz},\quad \sz= 4\pi\,.
\end{align}
However, after considering the corrections of $O(k^2)$, we need to pay careful attention  to the corrections of thermodynamical quantities.

\subsection{Thermodynamics and Linearised Hydrodynamics}
To study thermodynamics, we  come back  to  the quasi-static metric up to order $k^2$,
\begin{align}
\d {s}^2_{p+2} & =   g_{tt}(r) \d \ti^2 + g_{xx}(r)\de_{ij} \d {x}^i \d {x}^j  +  \frac{{2}}{\lz} \d{\ti}\d r , \nn \\
g_{tt}(r)&=-1-\frac{2\kz}{\lz^2}(r-\rc) + \frac{(r-\rc)}{\lz} \[  \Big(\frac{1}{4}\frac{ { r-\rc }}{  \rz-\rc }  - \frac{ p-1 }{p}\Big) \frac{\lz}{2 \kz}  - {\dzz} \] p k^2 ,\nn \\
g_{xx}(r) &=1 - \frac{(r-\rc)}{2\kz} k^2\, . \label{Rgtt}
\end{align}
We will consider thermodynamics  on the hypersurface $r=\rc$, and choose the ensemble in which the energy density is fixed with
the pressure  satisfying the Smarr relation, 
\begin{align}
{\tec}&\equiv\ec+  \zeta'_\ph (p k^2),\\
{\tpc}& \equiv \pc+ \zeta_\ph  (p  k^2)+\de\pc. \label{Rpk}
\end{align}
We choose the gauge $g_{tt}(\rz)=0$ in \eqref{Rgtt},
which keeps the position of the horizon fixed even with the corrections of $O(k^2)$,
then we obtain
\begin{align}
{{\dzz}}=  \(\frac{1}{p}-\frac{3}{4}  \) \frac{1}{\pc} ~\Rightarrow ~
g_{tt}(r) =-1-\frac{(r-\rc)}{\lz}   \({ \pc} - \frac{p}{4}\frac{k^2}{\pc} \frac{ r-\rz }{\rz-\rc} \) .
\end{align}
Under this choice, the temperature and energy density at $r=\rc$ associate with the metric \eqref{Rgtt} are  
\begin{align}
{\tTz} &= -\frac{\lz}{4\pi} g_{tt}'(\rz) 
=\frac{ \pc}{4 \pi}\(1-\frac{p}{4}\frac{k^2}{\pc^2}\), \label{RTk}\\
{\tsz} &= 4\pi \[g_{xx}(\rz)\]^{p/2} 
= 4\pi\(1 + \frac{p}{2}\frac{k^2}{\pc^2} \) +\Of. \label{Rsk}
\end{align}
According to the Smarr relation, the gauge parameter $\de\pc$ in \eqref{Rpk} reads
\begin{align}
{\tec}+ {\tpc} ={\tTz} {\tsz}~\Rightarrow ~\delta\pc=- \frac{k^2}{\pc}. 
\end{align}
Form the constraint equation \eqref{Rast}, in addition, we obtain
 \begin{align}
 D_a^\bot (\de{\pc}) = (\de {\ekz}) k^2 a_a +\Of,\qquad 
 \de  {\ekz} &= - \frac{1}{\pc}, \label{Rdtau}
 \end{align}
which will be used in the calculation of momentum relaxation rate.

{\bf Linearised Hydrodynamics.} ---
Under the procedures in \cite{Blake:2015epa},
we consider the linearised velocity and temperature field
\begin{align}
u^a\to (1, u_i(t)),\quad {\tTz}(t, x_i) \to {\tTz}+\delta {\tTz}(t, x_i).
\end{align}
In the linearised hydrodynamics, we only consider the linear perturbations of $u_i(t)$ and $\delta {\tTz}(t, x_i)$.
 Furthermore,  to study the frequency dependence of the transport coefficients,
it is enough to set $\p_i{{\tTz}}$ to be independent of the position $x^i$.
Then, the Ward identity caused by the momentum constraint \eqref{Rast} becomes
\begin{align}
\p_t \[(\ec_\zeta+\pc_\zeta)u_i\]  +\p_i \pc_\zeta&= - k^2   u_i +\frac{2 k^2}{\pc}   \p_t u_i +\cdots. \label{Rward}
\end{align}
Redefining the velocity $v_i$ such that it is proportional to the stress tensor $\langle {T^t}_i\rangle$ ,
\begin{align}
\langle {T^t}_i\rangle &=(\ec_\zeta+\pc_\zeta)u_i=({\tec}+{\tpc})v_i,\\
  v_i &=\frac{(\ec_\zeta+\pc_\zeta)}{({\tec}+{\tpc})}u_i  ,
\end{align}
the Ward identity \eqref{Rward} becomes
\begin{align}
\p_t \langle {T^t}_i\rangle +\p_i {\tpc}&= -\btz  \langle {T^t}_i\rangle-  {\ekz} k^2\p_t v_i +\cdots, \\
\btz &=\frac{ k^2  {\tsz}}{4\pi({\tec}+{\tpc}) } .  \label{RWard}
\end{align}
The coefficient correction on the right-hand side is
\begin{align}
 {\ekz} &= -\frac{2}{\pc} -\de{\ekz}=-\frac{1}{\pc} ,\\
\xi_0 &\equiv  \frac{{\ekz} \Tz }{\sz} =-1, \label{xiz}
\end{align}
and $\de{\ekz}$ comes from the contribution of $\de\pc$ in \eqref{Rdtau}.


Next, we will extract the thermodynamic response coefficient from the linearised hydrodynamics.
Assuming $\p_t v_i = -{\bi}\omega v_i  $, we have
\begin{align}
& v_i\[ k^2 \frac{{\tsz}}{4\pi}-{\bi}\omega \({\tec}+{\tpc}  -\frac{k^2}{\pc}  \) \]=  -{\tsz} \p_i {\tTz}  +\cdots, \end{align}
from which we obtain the solution of $v_i$
\begin{align}
& v_i =  - \frac{1}{1-{\bi}\omega\tauz}\frac{4\pi}{k^2} \p_i {\tTz}  +\cdots,
\end{align}
as well as the momentum relaxation rate
\begin{align} \tauz^{-1} = \btz\[1 + \frac{{\ekz} k^2}{ ({\tec}+{\tpc}) } \]^{-1}
= \frac{ k^2}{4\pi {\tTz} }\[1 - \xi_0 \frac{k^2}{\Tz^2}\] +O(k^6). \label{Rmrr}
\end{align}
The dimensionless number, ${{\ekz} \Tz }/{\sz} =-1$, was defined in \eqref{xiz}.
And from the definition of the  heat current, we can read off the heat current
\begin{align}
\langle Q_i \rangle\equiv \langle{T^{t}}_{i}\rangle=({\tec}+{\tpc}) v_i=- {\tka}_\omega\p_i {\tTz},
\end{align}
which leads to the  heat conductivity with momentum relaxation
\begin{align}
{\tka}_\omega&=\frac{1}{1-{\bi}\omega\tauz}\frac{4\pi {\tsz} {\tTz} }{k^2}, \\
  \tauz^{-1} &= \frac{ k^2}{4\pi {\tTz} }\[1 - \xi_0 \frac{k^2}{\Tz^2}\]+O(k^6),\label{tau0}
\end{align}
where ${\tsz} {\tTz}$ are given in \eqref{RTk} and  \eqref{Rsk}. 
In the DC limit $\omega\to 0$,
${\tka}_{\omega}$ reduces to the formulae
\begin{align}
{\tka}_{\DC}\equiv &\lim_{\omega\to 0} {\tka}_\omega = \frac{4\pi {\tsz} {\tTz} }{k^2}.
\end{align}
In  the next section, we will  confirm  these results from the near horizon limit of the cutoff AdS fluid.

\section{Momentum Relaxation in cutoff AdS Fluid}
\label{AdSCut}

In order  to relate our previous results on momentum relaxation in Rindler fluid with the momentum relaxation from fluid/gravity correspondence in AdS black brane \cite{Blake:2015epa},
in this section we start with Einstein-Hilbert action of $(p+2)$-dimensional AdS gravity with massless scalar fields 
 \begin{align}
S_\Lambda=\frac{1}{16\pi G_{p+2}}\int \d^{p+2} x\sqrt{-g} \[R+2\Lambda-\frac{1}{2}\sum_{\cA=1}^{p}  (\p \ph_\cA)^2 \]
-\frac{1}{8\pi G_{p+2}}\int \d^{p+1} x\sqrt{-\gamma} K.
\end{align}
The negative cosmological constant $\Lambda=-p(p+1)/2 \L^2$,
where $\L$ indicates the radius of AdS spacetime.
The equations of motion turn out to be
\begin{align}
R_{{\m}{\n}}-\frac{1}{2} R g_{{\m}{\n}}+\Lambda g_{{\m}{\n}} & = \frac{1}{2} \sum_{\cA=1}^p\[\p_\m\ph_\cA \p_\n\ph_\cA-\frac{1}{2} g_{\m\n}(\p \ph_\cA)^2\],\label{Aeomg}\\
\nabla^2\ph_\cA &= 0,\qquad \cA=1,2,...,p\,. \label{Aeomph}
\end{align}
There is an exact solution of the equations above, with the metric and scalar fields (see e.g. \cite{Andrade:2013gsa}),
\begin{align}
\d s^2_{p+2} &= \frac{r^2}{L^2}  \[ -\tf(r) \d {\tilde{t} \, }^2+\de_{ij} \d {x}^i \d {x}^j \] + \frac{L^2}{r^2} \frac{{\d} r^2}{\tf(r)}, \qquad \\
\tf(r) &= 1- \frac{\rh^{p+1}}{r^{p+1}}-\frac{L^2}{r^2} \frac{L^2 k^2}{2(p-1)}, \nn\\
\ph_{\cA} &=k x_i{\delta^i}_\cA, \qquad \cA=1,2,...,p . \label{Blake}
\end{align}

To study hydrodynamics with momentum relaxation, we will follow the set up  in \cite{Blake:2015epa} and 
consider $k$ as a small perturbation parameter, as well as identify $\p_a\sim k^2$.
Thus, in the following, the background metric we begin with is the black brane metric with an ingoing coordinate time $\ti$
\begin{align}
\d s^2 &= \frac{r^2}{L^2}  \[ -\frac{f(r)}{f(\rc)} \d \ti^2+\d \vec{x}^2 \] + \frac{2 }{\sqrt{f(\rc)}} \d \ti \d r ,
\qquad
f(r) =1- \frac{\rz^{p+1}}{r^{p+1}}.
\end{align}
The horizon is located at $r=\rz$.
The time coordinate $t$ has been rescaled as $\ti\to \sqrt{f(\rc)} \ti$ in order to keep the induced metric on the cutoff surface $r=\rc$ conformal flat, with $\gamma_{ab}=\frac{\rc^2}{L^2}\eta_{ab}$.
 The  temperature and entropy density associate with this metric turn out to be 
\begin{align}\label{Akappaf}
\Tf &=\frac{ \kf}{ 2\pi \lf},\qquad   \kf  \equiv \frac{\rz^2  {f'(r_0)}}{2 L^2}, \\
{\sfl} &= 4\pi \frac{\rz^p}{L^p}, \qquad      \lf  \equiv  \sqrt{f(\rc)} \,. 
\end{align}

\subsection{Fluid/Gravity Duality  in AdS Spacetime with a cutoff} 
After boosting the coordinates on the cutoff surface $\d t\to-u_a {\d} x^a, \de_{ij}\d{x^i}\d{x^j} \to h_{ab} {\d} x^a{\d} x^b$, and assuming the coordinates dependent of the horizon parameter $\rz(x^a)$ and velocity $u_a(x^a)$,
the metric and scalar fields which solve the equations of motion \eqref{Aeomg}  and \eqref{Aeomph} up to $O(k^3)$ turn out to be
\begin{align}\label{metriccutoffAdS}
\d {s}^2_{p+2} &={g}_{\mu\nu} \d x^\mu \d x^\nu= - \frac{2}{\lf} u_a \d x^a \d r+ {g}_{ab} \d x^a \d x^b,\\
{g}_{ab} &= {g}_{ab}^{{(0)}}+{g}_{ab}^{{(1)}} + O(k^4),\\
\ph_{\cA}&=\ph^{(0)}_{\cA}+\ph^{(1)}_{\cA}+\ph^{(2)}_{\cA}+ O(k^4). \label{Ascalar}
\end{align}
Then  zeroth order solution in derivatives expansion is
\begin{align}
{g}_{ab}^{{(0)}}&= - \frac{r^2}{L^2} \frac{f(r)}{f(\rc)}u_a u_b+  \frac{r^2}{L^2} h_{ab}, \quad h_{ab}\equiv \eta_{ab}+ u_a u_b,\label{Aback}\\  
\ph^{(0)}_{\cA} &=k x_i{\delta^i}_\cA, \qquad\quad \cA =1,2,...,p\,.
\end{align}
Notice that  the sources $\ph^{(0)}_{\cA} $ are chosen in the laboratory frame.
The first order solution of the metric ${g}_{ab}^{{(1)}}$ in derivative expansion can be
 decomposed, along the velocity $u_a$, into 
 \begin{align}
{g}_{ab}^{{(1)}}&=   g_{uu}^{(1)}(r) u_a u_b+ 2 u_{(a} g_{b)u}^{(1)}(r) 
+g_{hh}^{(1)}h_{ab}+F_\s(r)\sigma_{ab} +F_\ph(r)\sigma^\ph_{ab}.
\end{align}
After requiring the Dirichlet boundary condition at $r=\rc$ and regular boundary condition at the horizon of the black brane $r=\rz$,
the detailed formulas are solved as
\begin{align}
g_{uu}^{(1)}(r)&= \frac{ 1}{\lf^2 r^{p-1}} \int_{\rc}^r \d \tr{\tr^{p-2}}  \[ \frac{L^2{\tph}}{2 p} + 2 \tr \lf \theta
-  \frac{\tr }{\lf}  \(1-\frac{p-1}{2p}\frac{\rz^{p+1}}{\tr^{p+1}}\) \( 2\theta+  {\dzp}  \theta^\ph \) \]  , \nn\\
g_{au}^{(1)} (r)&=\frac{1}{\lf r^{p-1} } \int^r_{\rc} \d \tr \tr^p
\[ \frac{(a_a + D_a^\bot\ln\lf)- \djp a_a^\ph  }{\rc}
+\int^\tr_{\rc}\d\trp\(\frac{p(a_a-D_a^\bot\ln\lf)}{\trp^2}-\frac{F_1'(\trp)}{\trp}a_a^\ph \) \] ,\nn\\
g_{hh}^{(1)}(r)  &=  \frac{- r^{2}}{p\lf} \int^r_{\rc} \d \tr \frac{1}{\tr^2} \( 2\theta+{\dzp}  \theta^\ph\) ,\label{Ametric1}
 \end{align}
along with
\begin{align}
F_\s(r) &=- 2\lf r^2 \int_{\rc}^{r}\d\tr \frac{1}{\tr^2f(\tr)} \(1-\frac{\rz^p}{\tr^p}\), \nn\\
F_\ph(r) &=- \frac{ r^2}{p-1} \int_{\rc}^{r}\d\tr \frac{L^2}{\tr^3f(\tr)} \(1-\frac{\rz^{p-1}}{\tr^{p-1}}\).
\end{align}
Again, we have neglected the irrelevant terms $ (D\ph_{\cA}^{(0)})^2$.
In the solutions \eqref{Ametric1}, ${\dzp}$ depends on the gauge choice of the boundary fluid.
For the higher order solutions of the scalar fields in \eqref{Ascalar},
\begin{align}
\ph^{(1)}_{\cA} &
=  {(D\ph_{\cA}^{(0)})}F_1(r),~\quad F_1(r)=- {\lf} \int_{\rc}^{r}\d\tr \frac{L^2 }{\tr^2f(\tr)} \(1-\frac{\rz^p}{\tr^p}\).
\end{align}
And $\ph_{\cA}^{(2)}$ is solved as 
\begin{align} \label{Aphi2}
\ph_{\cA}^{(2)} = &  -\int_{\rc}^{r}\frac{{\d} {\tr} L^2}{\tr^{p+2} f(\tr)}\int_{\rz}^{\tr}{\d} {\trp}{\trp}^p
\Big\{ - (D \ph_{\cA}) \frac{ \lf^2}{{\trp}^p} {{\Big[}{\trp}^p F_1'({\trp})\Big(g_{uu}^{(1)}({\trp})- {\trp}\frac{ (p-1) }{(p+1)} \frac{2\lf^2 D\ln \lf}{(1-\lf^2)}\Big){\Big]}'} \nn\\
 &+ (D \ph_{\cA}){\Big[}\frac{L^2}{{\trp}^2}  \theta  +\lf F_1'({\trp})  \Big(\theta- \frac{ 2 p \lf^2 D\ln \lf }{(p+1)(1-\lf^2)}  \Big)
+\frac{\lf}{2}\(\frac{L^2}{{\trp}^2} +\frac{f({\trp})}{\lf} F_1'({\trp})\) {\trp}^2\Big(\frac{g_{hh}^{(1)}({\trp})}{{\trp}^2}\Big)' {\Big]} \nn\\
&+(D_a^\bot\ph_{\cA}){\Big[} ( a_b-D_b^\bot\ln \lf) + \frac{\lf}{{\trp}^{p-2 }} \({\trp}^{p-2 } g_{bu}^{(1) } \!({\trp})\)' {\Big]} \frac{L^2}{{\trp}^2} \eta^{ab} \nn\\
&+ \lf \Big( 2 F_1'({\trp})+  \frac{p}{r} F_1({\trp})\Big) {\Big[}(a^a D_a^\bot\ph_{\cA}^{(0)})  +\(D\ph_{\cA}^{(0)}  D\ln \lf\) \frac{(p+1)- (p-1) \lf^2 }{(p+1)(1-\lf^2)} {\Big]}\Big\} .
\end{align}
We have used the constraint equations in \eqref{Aeomg}, 
\begin{align}
\frac{\rc^2}{L^2}\[ \frac{f'(\rc)}{\lf } \theta - \frac{2p}{\rc} (D   \lf) \]
&=  \sum_{\cA=1}^p(D \ph_\cA^{(0)})  (D \ph_\cA^{(0)}) \frac{\rz^{p}}{\rc^{p}} +\Of , \\
\frac{\rc^2}{L^2}\[ \frac{f'(\rc)}{\lf } a_a + \frac{2p}{\rc} (D_a^{\bot}   \lf )+ D_a^{\bot}\Big( \frac{f'(\rc)}{\lf }\Big) \]
&= \sum_{\cA=1}^p(D_a^\bot \ph_{\cA}^{(0)}) (D \ph_{\cA}^{(0)})\frac{\rz^{p}}{\rc^{p}}  + \Of . \label{Amconstraint}
\end{align}

{\bf Dual Hydrodynamics.} ---
With these solutions, now we can calculate the dual stress tensor and scalar operators on the cutoff surface
\begin{align}
\langle{\cT^a}_b\rangle&=
-   \frac{\rc^{p+1}}{L^{p+1}}\[ 2\({K^a}_b-K {\g^a}_b \) + \frac{2p  {\cgL} }{L} {\g^a}_b+ \frac{\cp L}{p-1} \sum_{\cA=1}^p\Big((\p^a\ph_\cA\p_b\ph_\cA)-\frac{1}{2}{\g^a}_b (\p\ph_\cA)^2\Big)\]\Big|_{r=\rc}, \nn\\
\langle{\cO}\rangle&=  
- \frac{\rc^{p+1}}{L^{p+1}} \[n^{{\m}} \p_{{\m}} \ph_\cA+\frac{\cp L}{p-1}\p^2\ph_\cA\]\Big|_{r=\rc} .
\end{align}
After putting into the metric \eqref{metriccutoffAdS} and scalar fields \eqref{Ascalar}, the final results can be expressed as
\begin{align}
\langle{{\cT}_{ab}}\rangle  &= \cE  u_a u_b + \cP h_{ab} + \zeta_{\cc} \theta h_{ab} -2\eta_{\cc} \sigma_{ab} \nn \\ 
&+ \zeta'_\ph\theta^\ph u_a u_b+\zeta_\ph\theta^\ph h_{ab}+ 2 j_\ph a^\ph_{(a} u_{b)} -2\eta_\ph \sigma_{ab}^\ph+\Of, \label{ATab}  \\
\langle{\cO}\rangle & =
- (D \ph_{\cA}^{(0)})\frac{\rz^{p}}{L^{p}}
- \lf  \frac{\rc^{p+2}}{L^{p+2}}  \ph_{\cA}^{(2)'}\big|_{r=\rc} +\Of \label{AOI}.
\end{align}
The energy density $\cE$ and pressure $\cP$ could be read out from the  background metric \eqref{Aback}, 
\begin{align}
\cE &= \frac{\rc^{p+1}}{L^{p+1}}\[( {\cgL}-\lf)\frac{2p}{L}\], \\
\cP &= \frac{\rc^{p+1}}{L^{p+1}}\[(\lf- {\cgL})\frac{2p}{L} +\frac{\rc}{L} \frac{f'(\rc)}{\lf }\].
\end{align}
At the first order in derivative expansion, shear viscosity $\eta_{{\cc}}$ and bulk viscosity $\zeta_{{\cc}}$ are
\begin{align}
\eta_{{\cc}} &=\frac{\rc^p}{L^p}\[1+\frac{\lf}{2}F_\s'(\rc)\]=\frac{\rz^p}{L^p}, \quad ~~ \zeta_{{\cc}}=0.
\label{AShear}
\end{align}
As well as the coefficients which are contributed from the scalar fields,
\begin{align}
\eta_{\Ph}&=\frac{\rc^{p-1}}{L^{p-1}}\[\frac{\cp L}{2(p-1)}+\frac{\rc}{L}\frac{\lf}{2}F_\ph'(\rc)\]=\frac{L}{2(p-1)} \frac{ \rz^{p-1} -\rc^{p-1}(1- \cp\lf)}{\lf L^{p-1}}  , \\ 
 \zeta'_{\Ph} &= \frac{\rc^{p-1} }{L^{p-1}}  \[\frac{\rc}{L}{\dzp}  -  \frac{{\cp} L}{2(p-1)}\],\quad~~
 j_{\Ph} =\frac{\rc^{p-1} }{L^{p-1}}  \[\djp+ \frac{{\cp} L}{ (p-1)}\], \label{Ajphi}\\
 \zeta_{\Ph}&=\frac{\rc^{p-1} }{L^{p-1}}\[ \frac{{\cp} L}{2(p-1)} \frac{( p-2 )}{ p }-\frac{L}{2p\lf}+\frac{\rc}{L}\frac{\dzp}{2p} \(\frac{p+1}{\lf^2}-(p-1)\)
\] . 
\end{align}
At the second order in derivative expansion, the  term $\ph_{\cA}^{(2)'}\big|_{r=\rc} $ in $\OA$ at \eqref{AOI} could be read out from solution in~\eqref{Aphi2}.

The constraint equations  $\p_a\langle\cT^a_b\rangle=\p_a\ph_\cA \langle\cO_{\cA}\rangle$ turn  out to be
\begin{align}
 (\cE_\zeta+\cP_\zeta)\theta + u^b \partial_b \cE_\zeta
&  =  \sum_{\cA=1}^p (D \ph_\cA^{(0)}) \OA   , \\
 (\cE_\zeta+\cP_\zeta) a_a +{h_a}^b\p_b\cP_\zeta
&= \sum_{\cA=1}^p (D_a^\bot \ph_{\cA}^{(0)}) \OA  . \label{Aconsi}
\end{align}
And we have introduced the notations
\begin{align}
\cE_\zeta  &= \cE+ \zeta'_{\Ph}\theta^\ph,\qquad 
\cP_\zeta   =  \cP+ \zeta_{\Ph}\theta^\ph.
\end{align}
To meet with the Landau frame choice that $u_a\langle{\cT^a}_b\rangle=\cE_\zeta u_b$,
we can set $\djp= -  {{\cp} L}/{(p-1)} $ in \eqref{Ajphi}  such that $ j_{\Ph}$ in \eqref{ATab} vanishes.
Notice that the Smarr relation is satisfied on the cutoff surface
\begin{align}
{\cE} +{{\cP}} = \Tf {\sfl}=\frac{\rc^{p+2}}{L^{p+2}}\frac{f'(\rc)}{\lf},\quad   \Tf=\frac{\rz^2}{L^2} \frac{f'(r_0)}{4 \pi{\lf} }.
\end{align}
While, after considering the correction from momentum relaxation, we need to choose the ensemble in which $\cE_\zeta$ is fixed, as will be shown in the next subsection.

\subsection{Thermodynamics and Linearised Hydrodynamics}

We define the  thermodynamic quantities based on the following quasi-static metric, 
\begin{align}
\d {s}^2_{p+2} & =   g_{tt}(r) \d \ti^2 + g_{xx}(r) \de_{ij} \d {x}^i \d {x}^j +  \frac{{2}}{\lf} \d{\ti}\d r , \nn \\
g_{tt}(r)&=\frac{r^2}{L^2}\frac{f(r)}{\lf^2} +g_{tt}^{(1)}({r}),\qquad g_{xx}(r) =\frac{r^2}{L^2}  +g_{xx}^{(1)}({r}),\nn \\
g_{tt}^{(1)}({r}) &= \frac{ k^2}{\lf^2 r^{p-1}} \int_{\rc}^{r} \d \tr{\tr^{p-2}}  \[ \frac{L^2}{2  }  
-  \frac{ {p} \tr }{\lf}  \(1-\frac{p-1}{2p}\frac{\rz^{p+1}}{\tr^{p+1}}\)    {\dzp}  \], \nn\\
g_{xx}^{(1)}(r)&=  \frac{r^{2}  k^2}{ \lf}\(\frac{1}{r}-\frac{1}{\rc}\) {\dzp} .  
\end{align}
Notice that the new horizon $r=\rh$ satisfying $g_{tt}(\rh)=0$ is shifted as  
\begin{align} 
&g_{tt}(\rh)=0,\qquad  \rh \equiv \rz+\de\rz , \\
\Rightarrow~ &
\de\rz  =-\frac{g_{tt}^{(1)}({\rz})}{g_{tt}'({\rz})} = \frac{\lf}{4\pi} \frac{g_{tt}^{(1)}({\rz})}{ \Tf  }+\Of.
\end{align}
The local temperature and entropy density on the cutoff surface  are  given as
\begin{align}
{{\tTc}}&=- \frac{\lf}{4\pi} {g'_{tt}({\rh})} 
= \Tf\(\frac{\rz^{p}}{\rh^{p}}  +\frac{2\de\rz}{\rz} \)  - \frac{\lf}{4\pi} {g_{tt}^{(1)'}({\rz})}+\Of , \label{ATk} \\
{\tsc}&= 4\pi \[g_{xx}(\rh)\]^{p/2} =4\pi \frac{\rh^p}{L^p}\[1+ \frac{p L^2}{2 \rh^2}g_{xx}^{(1)}({\rh})\]+\Of  \label{Ask} .
\end{align}
We need to define the new energy density and pressure through
\begin{align}
{\tcE} &\equiv \cE+ \zeta'_{\Ph}(p  k^2) ,\qquad 
{\tcP} \equiv {\cP}+ \zeta_{\Ph} (p  k^2)+\de{\cP},
\end{align}
such that the Smarr relation is satisfied,
\begin{align} \label{SmarrAdS}
{\tcE}+ {\tcP} ={\tTc} {\tsc}  ~\Rightarrow ~
\de\cP = \frac{\rc^{p-1}}{L^{p-2}}  \frac{ k^2 }{  (p-1)}\[ \frac{1}{\lf} \(\frac{\rz^{p-1}}{\rc^{p-1}}- 1\)+\cp
 \] +\Of .
\end{align}
Interestingly, $\dzp$ does not appear in $\de\cP $ and 
\begin{align}
D_a^\bot (\de{\cP}) &=\frac{\rz^{p-1}}{L^{p-2}}  \frac{ k^2 }{\lf}
\[(D_a^\bot\ln \rz)  - \frac{D_a^\bot\ln\lf}{p-1}  \( 1 -\frac{ \rc^{p-1}}{ \rz^{p-1}} \)\]
= (\de{\ekf}) k^2 a_a,
\end{align}
where after using the constraint equation in \eqref{Amconstraint}, we can see that
 \begin{align}
 \de {\ekf} &=
 \frac{ -2\lf   }{(p+1) -(p-1)\lf^2} \[2-\frac{(p+1)}{(p-1)  \lf^2} \(  1- \frac{\rz^{p-1}}{ \rc^{p-1}} \)\frac{\rz^{2}}{\rc^{2}} \]\frac{\rz^{p-1}}{L^{p-2}}.
 \end{align}
 
{\bf Linearised Hydrodynamics.} ---
For the linearised hydrodynamics, again we consider the linearised velocity and the temperature field
\begin{align}
u^a\to (1, u_i(t)),\quad {\tTc} (t, x_i) \to  \tTc+\delta  \tTc(t, x_i).
\end{align}
The Ward identity yields the following momentum non-conservation equation \eqref{Aconsi}
\begin{align}
(\cE_\zeta+\cP_\zeta) \p_t u_i +\p_i \cP_\zeta&= -   \frac{k^2 {\tsc} }{4\pi} u_i -   ({\ekf} + {\de}{\ekf}) k^2 \p_t u_i +\cdots .
\end{align}
After  redefining the velocity $v_i$ such that
\begin{align}
\langle {\cT^t}_i\rangle &=(\cE_\zeta+\cP_\zeta)u_i=({\tcE}+{\tcP})v_i,\\
  v_i &\equiv \frac{({\cE}_\zeta+{\cP}_\zeta)}{({\tcE}+{\tcP})}u_i ,
\end{align}
the Ward identity for momentum non-conservation equation then up to order $O(k^4)$ becomes
\begin{align}
\p_t \langle {\cT^t}_i\rangle +\p_i {\tcP}&= - \bto \langle {\cT^t}_i\rangle  - {{\ekf}}{k^2}  \p_t v_i +  \cdots ,\\ 
\bto &=\frac{k^2 {\tsc}}{4\pi({\tcE}+{\tcP})} . \label{AWard}
\end{align}

Assuming $\p_t v_i = -{\bi}\omega v_i  $ and considering $\p_i {\tcP}= {\tsc}\p_i {\tTc}$,
which can be deduced from the first law of black hole thermodynamics  $\p_i {\tcE}= {\tTc}\p_i{\tsc}$ along with the Smarr relation in \eqref{SmarrAdS}, we then obtain
\begin{align}
& v_i\[ k^2 \frac{{\tsc}}{4\pi}-{\bi}\omega \({\tcE}+{\tcP}  +{\ekf}  k^2   \) \]=  -{\tsc} \p_i {\tTc}  +\cdots . \end{align}
From which we obtain the solution of $v_i$
\begin{align}
& v_i =  - \frac{1}{1-{\bi}\omega\tauo}\frac{4\pi}{k^2} \p_i {\tTc}  +\cdots,
\end{align}
as well as the momentum relaxation rate
\begin{align} \tauo^{-1} = \frac{{\tsc} k^2}{4\pi ({\tcE}+{\tcP})}\[1 + \frac{ \ekf  k^2}{  {\tcE}+{\tcP}  } \]^{-1} 
=\frac{k^2}{4\pi {\tTc} }\[1 - \xi_c \frac{k^2}{\Tf^2}\] +O(k^6). 
\end{align}
Here the coefficient $\ekf $ is given by 
\begin{align}
\ekf  =\lf \[\int_{\rz}^{\rc}\d \tr   \frac{\rz^{2} }{\tr^3f(\tr)}  \( 1-\frac{\rz^{p-1}}{\tr^{p-1}}\)  
-\frac{1 }{(p-1)\lf^2 } \frac{\rz^{2}}{\rc^{2}} \(  1- \frac{\rz^{p-1}}{\rc^{p-1}} \)\] \frac{\rz^{p-1}}{L^{p-2}} . \label{Ataok}
\end{align}

Thus, from the definition of the  heat current $\langle \cQ_i\rangle$, we can read off 
\begin{align}
\langle \cQ_i\rangle \equiv \langle {\cT^{t}}_{i}\rangle=({\tcE}+{\tcP}) v_i=-{\tka}_\omega\p_i {\tTc},
\end{align}
which leads to the  heat conductivity with momentum relaxation
\begin{align}\label{heatAdS}
{\tka}_\omega &=\frac{1}{1-{\bi}\omega \tauo}\frac{4\pi {\tsc} {\tTc} }{k^2}, \\
 \tauo^{-1}  &=\frac{k^2}{4\pi {\tTc} }\[1 -  \frac{{\ekf} \Tf }{\sfl} \frac{k^2}{\Tf^2}\] 
+O(k^6).\end{align}
In the DC limit $\omega\to 0$,
this expression reduces to the formulae in terms of the local entropy density ${\tsc}$ and temperature ${\tTc}$, which are given in \eqref{ATk} and \eqref{Ask},
\begin{align}
{\tka}_{\DC} &\equiv \lim_{\omega\to 0}{\tka}_\omega = \frac{4\pi {\tsc} {\tTc} }{k^2}.
\end{align}
For simplification,
we can rewrite  $\ekf$ in \eqref{Ataok} as the dimensionless form
\begin{align}
\xi_c\equiv &\frac{{\ekf} \Tf }{\sfl}  = (p+1 ) \[\Ft_p(\rc)
-\frac{\rc \Ft'_p(\rc)}{(p-1) } \], \label{xic} \\
\Ft_p(r)\equiv &\int_{\rz}^{r} \frac{\d \tr  \,\rz^{2} }{\tr^3f(\tr)}  \( 1-\frac{\rz^{p-1}}{\tr^{p-1}}\). \label{AtaokTs}
\end{align}

\section{From Conformal Fluid to Rindler Fluid} 
\label{RGflow}
In the fluid/gravity duality with a finite cutoff surface, 
the running of the cutoff surface is interpreted as the holographic Wilson renormalization group flow~\cite{Bredberg:2010ky,Cai:2011xv,Kuperstein:2011fn,Brattan:2011my}, a recent discussion of the dual field theory on the finite cutoff surface can be found in \cite{McGough:2016lol}. However, it has been found that the first order transport coefficients, such as the ratio of shear viscosity over entropy density $\eta_c/s_c=1/4\pi$, does not run with the cutoff surface. In the following, we will show that the dimensionless  sub-leading correction $\xi_c$, which is defined in \eqref{xic},
will run along with the cutoff scale $r_c$.

\begin{figure}[h] 
\begin{center}
\includegraphics[width=11cm]{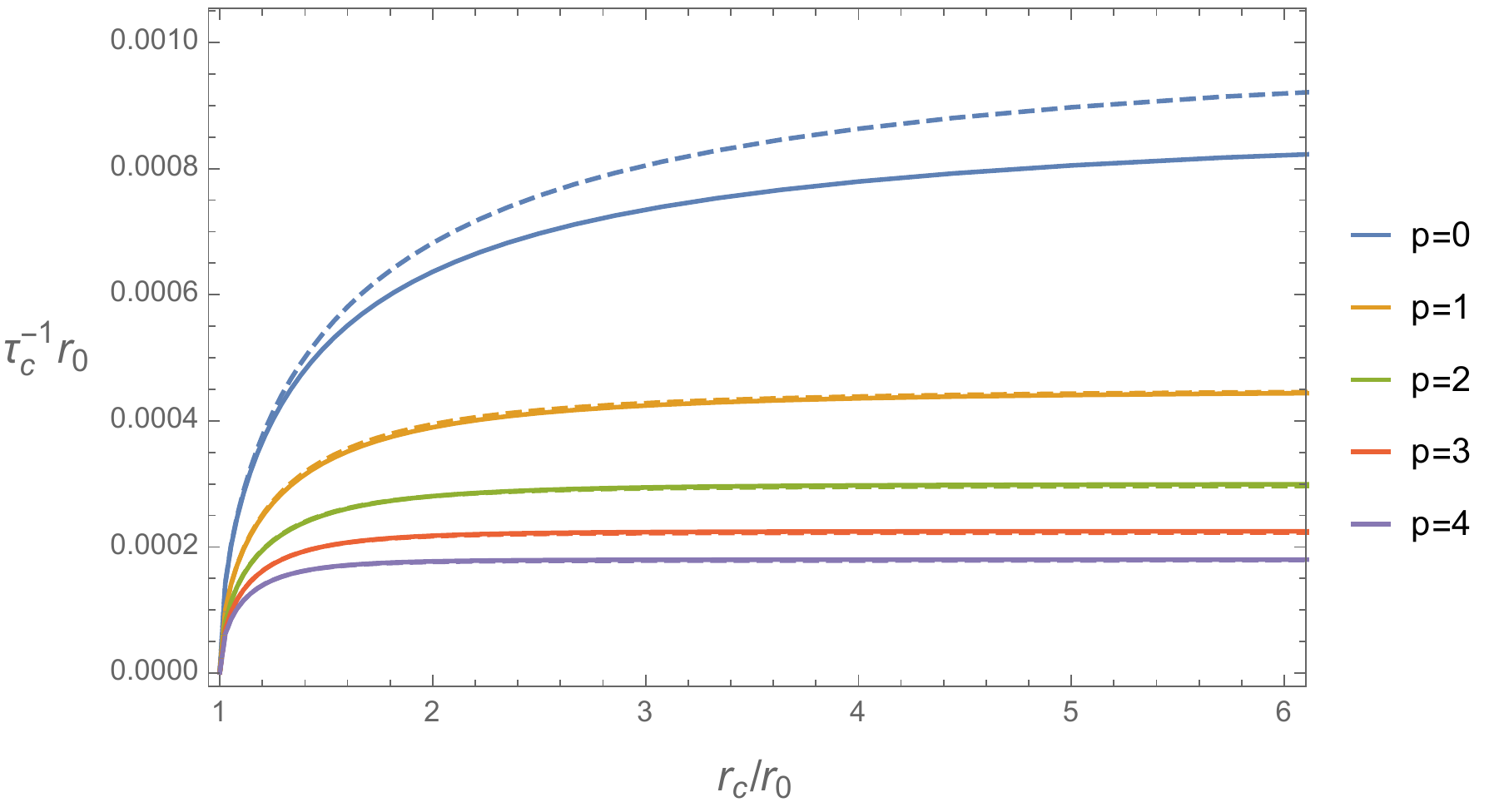}
\caption{The momentum relaxation rate $\tauo^{-1}$ in terms of the position of the cutoff surface $r_c$.
We use $\rz$ to normalize the unite and take $k=0.03$. The holographic fluids live in the $p+1$ dimensional spacetime, and $p=0,1,...,4$ from top to down.
The solid lines indicate the leading order contribution of $\tauo^{-1}$ in \eqref{tauc1}, up to $k^2$.
In the dashed lines the sub-leading terms in $\tauo^{-1}$  up to $k^4$ have been included.
}
\label{RG1}
\end{center}
\end{figure}

\begin{figure}[h] 
\begin{center}
\includegraphics[width=11cm]{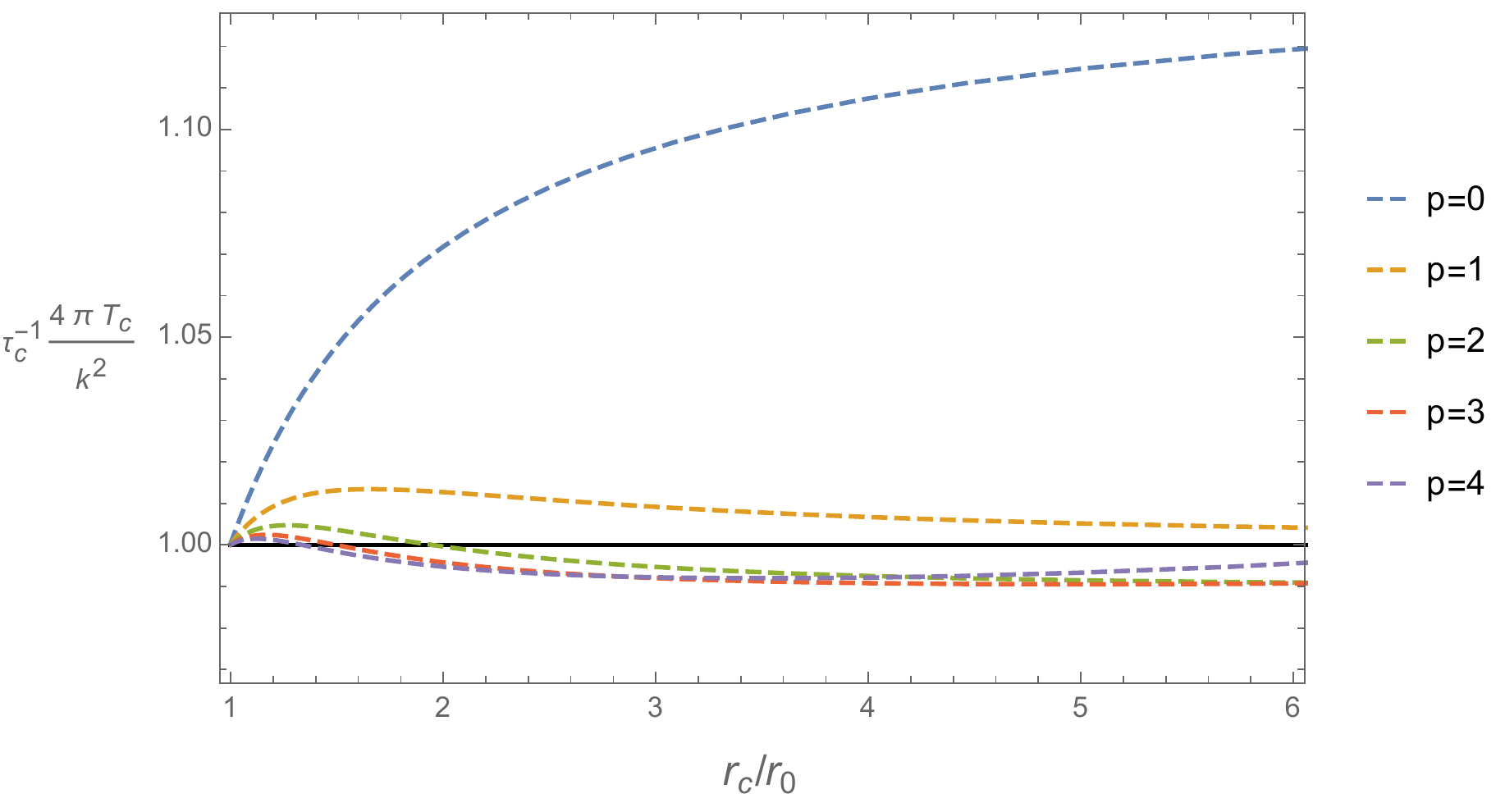}
\caption{The momentum relaxation rate $\tauo^{-1}$, which is multiplied by $\frac{4\pi T_c}{k^2}$, in terms of the position of the cutoff surface $r_c$. The black solid lines indicate that at the leading order the formula $\tauo^{-1} \frac{4\pi T_c}{k^2}$ is independent of the cutoff scale $r_c/r_0$.
However, the dashed lines indicate the $r_c$ dependence of $\tauo^{-1} \frac{4\pi T_c}{k^2}$, with the sub-leading contributions in \eqref{tauc1}.
Roughly, from top to down  $p=0,1,...,4$.}
\label{RG12}
\end{center}
\end{figure}

\begin{figure}[h] 
\begin{center}
\qquad\includegraphics[width=11cm]{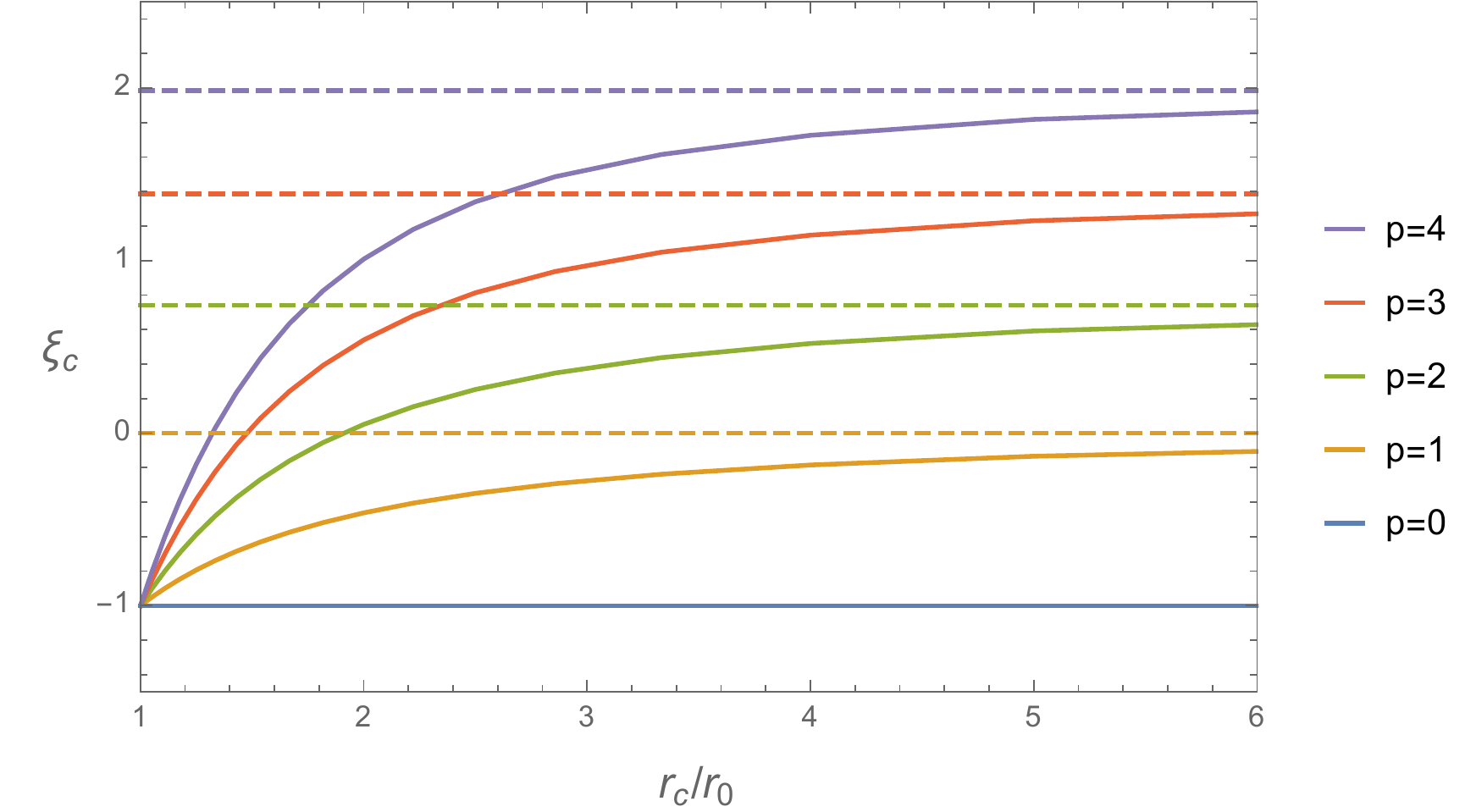}
\caption{The dimensionless coefficient $\xi_c$ in terms of the position of the cutoff surface $r_c/r_0$.
The holographic fluids live in the $p+1$ dimensional spacetime, and $p=0,1,...,4$ from the bottom up.
$\xi_c$ is the key coefficient to indicate the sub-leading correction to the momentum relaxation rate in \eqref{tauc1}.
The solid lines are the direct plots of $\xi_c$ in \eqref{xic}, and the dashed lines are taken from $\xi_\infty$ in \eqref{xif}.
For the case of $p=0$ and $p=1$, $\xi_c$ is always negative. They agree with the plots in Figure \ref{RG12} after considering \eqref{tauc1}.
 }
\label{RG2}
\end{center}
\end{figure}

The breaking of translational invariance modifies the conservation equations of relativistic hydrodynamics into
$ \p_a {\cT^a}_b=\p_b \ph_\cA\langle{\cO_\cA}\rangle $,
where the Ward identity for the stress tensor controls how momentum relaxes to equilibrium through scattering of the scalars.
Notice that beyond the leading order that was studied in  \cite{Hartnoll:2007ih} with $ \p_a  {\cT^a}_i =- \tauo^{-1}  {\cT^a}_i $,
the new holographic Ward identity up to {order $k^4$}   suggested in \cite{Blake:2015epa} is
\begin{align}
 \p_t   {\cT^t}_ i +\p_i \cP & =- \bto   \cQ_i  -{\ekf} k^2 a_i ,\qquad \bto=\frac{k^2 }{4\pi {\tTc} },
\end{align} 
with the acceleration $a_i=\p_t v_i$. 
It is in \eqref{RWard} for our  Rindler fluid, and in \eqref{AWard} for our cutoff AdS fluid.
For the cutoff AdS fluid, the 
momentum relaxation rate up to {order $k^4$} are
 \begin{align}
\tauo^{-1}  &=\frac{ k^2 }{4\pi {\tTc}}\(1-\xi_c \frac{ k^2 }{ \tTc^2 }  \) , \qquad \xi_c=  \frac{{\ekf} \Tf }{\sfl}.\label{tauc1}
\end{align}
The value of ${\xi_c}$ is given in \eqref{xic}, which is one of our main conclusions.
In the following, we will take both of the near horizon limit and near boundary limit, and plot the running of relaxation rate $\tauo^{-1}$(Figure \ref{RG1} and \ref{RG12}) and sub-leading coefficient ${\xi_c}$(Figure \ref{RG2})  along with the cutoff surface $\rc$.

{\bf Near horizon limit}. --- 
In order to take the near horizon limit $\rc\to\rz$, 
and match with the gauge choice in the Rindler fluid,
we can choose the gauge $g_{uu}^{(1)}(\rz)=0$ in \eqref{Ametric1} and fix  ${\dzp}$ through   
\begin{align}\label{HorizonGauge}
 g_{uu}^{(1)}(\rz) =0 ~\Rightarrow~ \dzp=\frac{L^2\lf}{\rz(p+1)}\(1-\frac{p}{2}\frac{\rc-\rz}{\rz}+...\).
\end{align}
We need to make the coordinate transformation
\begin{align}
x^a \to \frac{\rc}{L} x^a,\qquad \frac{\rc}{L}  \lf  \to \frac{\rz}{L}  \sqrt{ f'(\rz) (\rc-\rz)}   =  \lz. \label{Aconformal}
\end{align}
The near horizon limit indicates 
\begin{align}
\frac{f(r)}{f(\rc)} &\to \frac{f'(\rz)(r -\rz)}{f'(\rz)(\rc- r_0)} +O(\lf^2).
\end{align}
After identifying
\begin{align}
 2\kf  &=\frac{\rz^2}{L^2}f'(\rz) = 2\kz,
\end{align}
such that $\Tf\to \Tz$,
we can recover the Rindler fluid with momentum relaxation.
In particular, the following dimensionless quantity in \eqref{xiz} is re-obtained from the near horizon limit,
\begin{align}\label{xiz4}
\lim_{\rc\to\rz} \xi_c = \xi_0=-1 .
\end{align}
Then from $\tauo^{-1}$ in \eqref{tauc1},  we can also recover the formula of $\tauz^{-1}$ in \eqref{tau0}.
Notice that in order to keep the correct physical dimensions, we have restored the surface gravity $\kz$ in Rindler fluid instead of setting $2\kz=1$ in the literature  \cite{Compere:2011dx}, and we keep the AdS radius $L$ in the cutoff  AdS fluid. After changing into the notations of the conformal coordinates with \eqref{Aconformal},
we can also recover the conversion and results in \cite{Pinzani-Fokeeva:2014cka}.

{\bf Near boundary limit.} ---
The near boundary limit  $\rc\to\infty$ of the cutoff surface in AdS is   easier to reach, since we  kept the conformal factor in the metric \eqref{Aback}.
Refer to the procedure in \cite{Pinzani-Fokeeva:2014cka},
we can simply set
\begin{align}
\cgL \to 1,\quad {\cp}\to 1, \quad \lf\to 1,
\end{align}
to recover all results at the AdS boundary.
In particular, the dimensionless number
\begin{align}\label{xif}
&\lim_{\rc\to\infty} \xi_c  = \xi_\infty   \equiv(p+1) \Ft_p(\infty),\\
 &\Ft_p(\infty)  \equiv \int_{\rz}^{\infty} \frac{\d \tr  \,\rz^{2} }{\tr^3f(\tr)}  \( 1-\frac{\rz^{p-1}}{\tr^{p-1}}\).\label{xiinfty}
\end{align}
For example, $ \Ft_2(\infty)=(9\ln 3-\sqrt{3}\pi)/18$ and $ \Ft_3(\infty)= {\ln 2}/{2}$ match with the values in
\cite{Blake:2015epa,Blake:2015hxa}.
Intriguingly, the factor $\Ft_p(\infty)$ given in \eqref{xiinfty} also appears in the second order transport coefficients of the holographic conformal fluid~\cite{Bhattacharyya:2008mz}. It would be interesting to study the second order hydrodynamics with momentum relaxation in Rindler fluid and cutoff AdS fluid.

In Figure \ref{RG1}, we plot the momentum relaxation rate $\tauo^{-1}$ in terms of the position of the cutoff surface $r_c$. We have set $\rz=1$, $k=0.03$ and use $\rz$ to normalize the units. The holographic fluids live in the $p+1$ dimensional spacetime, and $p=0,1,...,4$ from top to down. The solid lines indicate the leading order contribution of $\tauo^{-1}$ in \eqref{tauc1} up to $k^2$. In the dashed lines  the sub-leading terms in $\tauo^{-1}$ up to $k^4$ have been included.
In order to understand the meaning of the vanishing of momentum relaxation rate $\tauo^{-1}$ in near the horizon limit,
it would be more obvious to write down the leading order contribution in terms of $k^2$ that
\begin{align}\label{tauc}
\tauo^{-1}  &=\frac{k^2}{4\pi {T_c} } =\frac{k^2}{4\pi T_0}\frac{1}{\sqrt{f(r_c)}}, \qquad
T_0 \equiv \frac{\rz^2  {f'(r_0)}}{4\pi L^2}.
\end{align}
As in the near horizon limit $f(r_c)\to 0$,  the local temperature $T_c=\frac{T_0}{\sqrt{f(r_c)}}$ will be divergent due to the Tolman relation, 
which lead to the vanishing of momentum relaxation rate $\tauo^{-1}$.
While near the boundary limit $f(r_c)\to1$, and $\tauo^{-1}$ approach a finite value at each dimension.

Actually in the holographic models, the temperature and the Wilson RG scale are two independent parameters.
In our case, it is not necessary to identify the temperature $T_c$ with the holographic Wilson RG scale $r_c$.
Through taking a finite cutoff $r_c$, the dual fluid on the cutoff surface has been deformed from ``conformal fluid'' into ``cutoff AdS fluid''.
From the action formula $S_{\text{cutoff}}=S_{\text{CFT}} -  S_{\text{AdS}}|_{r_c}^{\infty}$,
it is more natural to explain $r_c$ as the UV cutoff of the momentum scale.
In Figure \ref{RG12}, we show that at leading order the momentum relaxation rate $\tau_c^{-1} \equiv  \frac{k^2}{4 \pi T_c}$,
which means that if fixing the temperature $T_c$ and relaxation strength $k$,  the relaxation rate $\tau_c^{-1}$ is independent of the UV cutoff $r_c$.
 It is not surprise because we only  consider weak strength $k$ and hydrodynamic limit with small wave number.
Thus, the effective  ``cutoff AdS fluid'' shares the same IR physics of the conformal fluid such as the momentum relaxation rate  $\tau_c^{-1} = \frac{k^2}{4 \pi T_c}$, the shear viscosity $\eta_c=s_c/4\pi$, and others. In this sense, Rindler fluid can also be considered as an effective theory of the conformal fluid after taking $r_c\to r_0$.
However, in Figure \ref{RG12}, the dashed lines indicate the $r_c$ dependence of $\tauo^{-1} \frac{4\pi T_c}{k^2}$, with the sub-leading contributions in \eqref{tauc1}.
Roughly, from top to down  $p=0,1,...,4$. To see their tendency more clearly, we need to plot the the dimensionless coefficient $\xi_c$  in \eqref{tauc1}.

In Figure \ref{RG2}, we plot the dimensionless coefficient $\xi_c$ in terms of the position of cutoff surface $r_c$.
The solid lines are the direct plots of $\xi_c$ in \eqref{xic}, and the dashed lines are taken from $\xi_\infty$ in \eqref{xif}.
In summary, the dimensionless constant in \eqref{xic} has two limits,
\begin{align}\label{xiRG}
\lim_{\rc\to\,\rz} \xi_c  &=\xi_0=-1,\\
\lim_{\rc\to\,\infty} \xi_c &=\xi_\infty=(p+1) \Ft_p(\infty).
\end{align}
Near the horizon, it recovers the correction in Rindler fluid \eqref{xiz}, 
and near the infinity boundary, it recovers the correction in AdS fluid \cite{Blake:2015epa,Blake:2015hxa}.
Thus, we have shown that even with the weak momentum relaxation,
the Rindler fluid can be still considered as the IR fixed point of holographic Wilson RG Flow in fluid/gravity duality,
and the conformal fluid at the boundary would plays the role as the UV fixed point.
On the other hand, although the Rindler fluid and conformal fluid share the same $\eta/s=1/4\pi$,
the sub-leading coefficient $\xi_0=-1$ in \eqref{xiz} and $\xi_\infty=(p+1) \Ft_p(\infty)$ in \eqref{xif} are different from each other.
 Also,  in Figure \ref{RG2}, we have shown that how they related with each other along with the running of a cutoff surface.



\section{Conclusion}
\label{conclusion}
In this paper, we first introduced the weak momentum relaxation into Rindler fluid,
which lives on the timelike cutoff surface in Rindler frame.
The translational invariance is broken by massless scalar fields with weak strength $k$.  
Additionally,  the order of derivative expansion in the relativistic fluid is assumed to be $\p_a\sim k^2$.
We then solved the gravitational field and scalar field equations up to {order $k^3$},
and obtained the heat conductivity of Rindler fluid.
From the Ward identity up to {order $k^4$}, we obtained the momentum relaxation rate up to {order $k^4$} in \eqref{Rmrr}.
Through introducing a finite cutoff in AdS spacetime and considering both of the near horizon limit and near boundary limit,
we also showed that how the momentum relaxation in Rindler fluid flows to the dual fluid living on the boundary of AdS.
In particular, we obtain the dimensionless coefficient ${\xi_c}$  in \eqref{xic}.
In the IR limit that $\rc\to\rz$,  we have $\xi_c\to\xi_0$ in \eqref{xiz}.
And in the UV limit that $\rc\to\infty$, we have $\xi_c\to\xi_\infty$ in \eqref{xif}.
It would be interesting to study the physical meaning of $\xi_c$ in the further studies.

Our setup is  originated from the ansatz in \cite{Bredberg:2010ky}, 
where the effective fluid dual to the gravity inside a cutoff $r_c$ has been studied. 
Since the metric region from $r_c$ to the  AdS boundary is all removed, the $r_c$ dependence of physical quantities, such as the diffusion coefficient, shear viscosity, is interpreted as Wilson RG flow in the dual fluid.
With the helpful and technical studies in \cite{Pinzani-Fokeeva:2014cka},
we have improved the induced metric on the cutoff as $\gamma_{ab}=\frac{r_c^2}{L^2} \eta_{ab}$, instead of $\gamma_{ab}= \eta_{ab}$ in \cite{Bredberg:2010ky}.
The conformal factor $\frac{r_c^2}{L^2}$ is kept such that  the dual conformal fluid at the AdS boundary can be  easily reached through taking $r_c\to \infty$.
Recent progress of the dual theory on the finite cutoff surface in AdS$_3$ appears in \cite{McGough:2016lol}, where the ``cutoff AdS theory''  is found to be exactly matched with the $T\bar{T}$ deformation of the CFT$_2$ at the boundary.
Although the higher dimensional generalization is still unclear,
 the dual ``cutoff AdS theory'' can be considered as either the deformation of CFT, or a kind of effective field theory, such as  
$S_{\text{cutoff}}=S_{\text{CFT}} -  S_{\text{AdS}}|_{r_c}^{\infty}$.
In this sense,  the action of Rindler fluid can be formally written as $S_{\text{Rindler}}=S_{\text{CFT}} - S_{\text{AdS}}|_{r_0+\epsilon}^{\infty}$,
with $r_0$ the position of the horizon.
In this paper, for the fluid living on the finite cutoff surface in AdS spacetime with momentum relaxation, we have checked that we can recover the Rindler fluid in the near horizon limit, and we can also recover the boundary fluid dual to AdS in the asymptotic infinity limit. 

What's more, it would be more interesting to consider the charged fluid in further work.
In the holographic condensed matters, the DC response coefficient  becomes infinite when the system is translational invariance.
In order to mimic the properties of lattice structure in real materials, there are extensive studies incorporating the impurity into the system by means of momentum relaxation. 
Moreover, the holographic transport coefficients should be well defined at small frequency limit. 
We can categorize the models into two types of studies depending on the  conditions at AdS boundary as below.
One is the inhomogenous boundary condition, where the dual boundary field is inhomogeneous in the spatial direction so that source becomes spatially dependent and partial differential equations have to be solved.  
See, for example, the lattice models in \cite{Horowitz:2012ky,Horowitz:2012gs, Donos:2013eha}.
The other one is the homogenous boundary condition, 
which provides an alternative way to dissipate momentum, and one can avoid the subtlety of coordinate dependent stress  tensor.  In particular, by breaking the diffeomorphism invariance leading to non-divergent free stress tensor.
For example, in holographic massive gravity 
(see e.g. \cite{Vegh:2013sk,Davison:2013jba,Blake:2013bqa,Blake:2013owa}),
the finite transport coefficients with momentum relaxation are observed  from holographic point of view. 
What's more, in the simple momentum relaxation model that been used in our paper, 
 scalar fields are chosen in such a way that bulk solutions are homogeneous and isotropic.
It is also feasible to generalise the approach to anisotropic case caused by different momentum relaxations \cite{Khimphun:2016ikw,Khimphun:2017mqb}, or spherical fluid dual to the black holes in massive gravity  \cite{Cai:2014znn,Park:2016slj}, which will be included in our further studies.

\allowdisplaybreaks

\begin{center} {\bf Acknowledgments}\\ \end{center}
This work is supported by APCTP at Pohang, CQUeST at Sogang University, through National Research Foundation of Korea (NRF),
Korea Ministry of Education, Science and Technology, Gyeongsangbuk-Do and Pohang City.  S. Khimphun and B.\,-H.\,Lee was partly supported by Sogang University Research Grant (No.\,201619067.01),   NRF grant(No.\,2014R1A2A1A01002306)(ERND)  funded by MSIP. C. Park was partly supported by Basic Science Research Program through NRF grant(No.\,2016R1D1A1B03932371)  funded by the Ministry of Education. Y.\,-L.\,Zhang was partly supported by YST program at APCTP.
We thank the referees' comments on the RG flow, which motivate our figures and discussions.

\end{document}